# Coherent and incoherent superposition of transition matrix elements of the squeezing operator.

**Sándor Varró**[1,2]

1) Wigner Research Centre for Physics, Eötvös Loránd Research Network, Budapest, Hungary

2) ELI-ALPS Attosecond Light Pulse Source, ELI-HU Non-Profit Ltd., Szeged, Hungary

**Abstract.** We discuss the general matrix elements of the squeezing operator between number eigenstates of a harmonic oscillator (which may also represent a quantized mode of the electromagnetic radiation). These matrix elements have first been used by Popov and Perelomov (1969) long ago, in their thorough analysis of the parametric excitation of harmonic oscillators. They expressed the matrix elements in terms of transcendental functions, the associated Legendre functions. In the present paper we will show that these matrix elements can also be derived in a different form, expressed by the classical Gegenbauer polynomials. This new expression makes it possible to determine coherent and incoherent superpositions of these matrix elements in closed analytic forms. As an application, we describe multiphoton transitions in the system "charged particle + electromagnetic radiation", induced by a (strong) coherent field or by a black-body radiation component (with a Planck-Bose photon number distribution). The exact results are compared with the semi-classical ones. We will show that in case of interaction with a thermal field, the semi-classical result (with a Gaussian stochastic field amplitude) yields an acceptable approximation only in the Rayleigh-Jeans limit, however, in the Wien limit it completely fails.

**1. Introduction**

The „non-classical states of light", in particular, „squeezed light" have been a subject of intensive theoretical and experimental research for many decades (Stoler 1970, 1971, Yuen 1976, Walls 1979, Loudon and Knight 1987, Dodonov 2002, Dodonov and Man'ko 2003, Andersen et al 2016), and have become standard subjects in any textbooks on quantum optics (Scully and Zubairy 1997, Loudon 2000, Schleich 2001). The phenomenon „squeezing of the amplitudes" (of a mechanical oscillator, or of a component of electromagnetic radiation) often appears in parametric processes (Husimi 1953, Louisell, Yariv and Siegman 1961, Mollow and Glauber 1967, Popov and Perelomov 1969, Malkin, Man'ko and Trifonov 1970, Milburn and Walls 1983, Kiss, Janszky and Ádám 1994, Dodonov 2003, Fedorov *et al* 2008, Straupe *et al* 2011). Concerning the oscillatory behaviour of the photon number distribution in highly-squeezed coherent states, see Wheeler (1985), Schleich and Wheeler (1987), Schleich, Walls and Wheeler (1988), Schleich (2001). The squeezing operator is the generator of the Bogoljubov transformation, which has played an important role in the theory of superconductivity (Bogoljubov 1958, Valatin 1958, Dodonov 2007). The quantum dynamics of charged particles in time-varying external (electro)magnetic fields also shows interesting phenomena, related to non-classical states of linear oscillators (Malkin, Man'ko and Trifonov 1970, Varró 1984, Varró and Ehlotzky 1985, 1987, Dodonov 2018, Dodonov and Horovits 2021). Recently the interest in entangled states is more pronounced then in nonclassical pure states. Anyway, undoubtedly, the „classic non-classical states" are the squeezed coherent states, which have been the ones most extensively studied in the last decades. It is lesser known by the quantum optics community that such states of the radiation field naturally appear as exact stationary solutions in the simplest (and most fundamental) system of quantum electrodynamics, namely in the interaction of a free electron with a quantized radiation field (see Berson 1969, Fedorov and Kazakov 1973, Bergou and Varró 1981a-b, Gazazyan and Sherman 1989, Varró 2021a). In the theory of intereactions of strong laser radiation with electrons such exact solutions of the Schrödinger or Dirac equations are of outstanding importance in treating the multiphoton processes non-perturbatively (though the laser field is mostly taken as an external field).

Our motivation for the present study has been to calculate the coherent and incoherent superposition of multiphoton transition amplitudes, which we have derived in the frame of non-relativistic quantum electrodynamics, for considering high-order harmonic generation (Varró 2021a). In the non-perturbative treatment of such processes one neccesarily encounters matrix elements between photon number eigenstates of the displacement and squeezing operators (which are standard objects in quantum optics). It is very remarkable that, though the photon number distributions of squeezed coherent states (Hermite polynomials), or displaced number states (Laguerre polynomials) have long been known, the matrix elements $S_{mn}=\langle m|S|n\rangle$ of the squeezing operator alone have not been published in a classical polynomial form. More precisely, $S_{mn}$ has been expressed by Bargmann (1947) in his paper on the irreducible representation of the Lorentz group, in terms of hypergeometric functions (see also Sannikov 1965). In the context of parametric processes, Popov and Perelomov (1969) and Malkin, Man'ko and Trifonov (1970) derived a form of $S_{mn}$, expressed in terms of associated Legendre functions. Later $S_{mn}$ has been related also to Jacobi polynomials (see, in particular, the very thorough recent studies of Wünsche 2017a-b), however a compact, classical polynomial expression has not been derived (see also Tanabe 1973, Rashid 1975, Satyanarayana 1985 and Mendaš and Popović 1995). Recently we have derived a compact classical polynomial expression, in term of Gegenbauer polynomials (Varró 2021b), and the larger part of the present paper will be devoted to the derivation and the main properties of the matrix element $S_{mn}$ of the squeezing operator.

In Section 2 we summarize the main steps of the derivation of the Gegenbauer polynomial expression of the matrix elements of the squeezing operator between photon number eigenstates. In order to have the paper self-contained, we have included an Appendix which contains all the necessary details of this derivation. We shall also present some numerical illustrations to display the main fetures of the photon number distribution in a squeezed number state. In Section 3 we apply the new formula for the matrix elements for calculating the coherent and incoheren superpositions of multiphoton squeezing transitions.

## 2. The matrix elements of the squeezing operator between photon number eigenstates.

In the present section we summarize the basic steps of the calculation of the matrix elements between the number eigenstates of the squeezing operator, $S_{m,n} = \langle m|S(\xi)|n\rangle$, where

$$S(\xi) = \exp[\tfrac{1}{2}\xi(a^+)^2 - \tfrac{1}{2}\xi^*(a)^2], \quad S(\xi) = \exp(\xi K_+ - \xi^* K_-), \quad \xi = |\xi|e^{i\varphi}, \qquad (1)$$

with $K_+ = \tfrac{1}{2}(a^+)^2$ and $K_- = \tfrac{1}{2}(a)^2$. Here $a$ and $a^+$ are the photon absorption and emission operators, respectively. (In a more general context, they are the quantum amplitudes associated to the decrease and increase by 1 of the excitation index of the system, like a mechanical linear oscillator.) The $a$ and $a^+$ satisfy the commutation relation $[a, a^+] = aa^+ - a^+a = I$, where $I$ is the unit operator of the complex Hilbert space $\mathcal{H}$, which models the system. The effects of $a$ and $a^+$ on the Fock states $|n\rangle$ (which are the photon number eigenstates, satisfying the eigenvalue equation $a^+a|n\rangle = n|n\rangle$) are expressed by equations $a|n\rangle = \sqrt{n}|n-1\rangle$ and $a^+|n\rangle = \sqrt{n+1}|n+1\rangle$. For the ground state $a|0\rangle = \vec{0}$, where $\vec{0}$ is the zero vector of $\mathscr{H}$. The Fock states $|n\rangle$ form a complete orthogonal set ($\langle m|n\rangle = \delta_{m,n}$) in $\mathscr{H}$. By introducing $K_0 = \frac{1}{4}(aa^+ + a^+a)$, we have the closed set of generators of a special Lie algebra of the $SU(1,1)$ group, having the commutation relations $[K_0, K_\pm] = \pm K_\pm$, and $[K_-, K_+] = 2K_0$. In Appendix A we summarize some basic properties of this group, on the basis of which we derive the normally ordered form of $S(\xi)$. The details of the derivations are presented in Appendix A, in the present section we just outline the basic steps, leading to the final result.

According to Eq. (A.5), the normal form of $S(\xi)$ becomes

$$S(\xi) = \exp[\tfrac{1}{2}\zeta(a^+)^2]\exp[-\eta\tfrac{1}{2}(a^+a+\tfrac{1}{2})]\exp[-\tfrac{1}{2}\zeta^*(a)^2], \quad \eta = 2\log\cosh|\xi|, \quad \zeta = e^{i\varphi}\tanh|\xi|. \tag{2}$$

By consecutively applying the relation $a^2|n\rangle = \sqrt{n(n-1)}|n-2\rangle$, we have the finite sum

$$\exp(-\tfrac{1}{2}\zeta^*a^2)|n\rangle = \sum_{k=0}^{[n/2]} \frac{(-\zeta^*/2)^k}{k!}\sqrt{n(n-1)(n-2)\cdots(n-2k+1)}|n-2k\rangle, \tag{3}$$

as the right-hand factor in the scalar product $S_{m,n} = \langle m|S(\xi)|n\rangle$. Similarly, the left-hand factor becomes also a finite sum, running up to $[m/2]$. The operator in the middle of the normal form in Eq. (2) results in a multiplication for any $|k\rangle$ by a number, namely, $\exp[-\eta\tfrac{1}{2}(a^+a+\tfrac{1}{2})]|k\rangle = \exp[-\eta\tfrac{1}{2}(k+\tfrac{1}{2})]|k\rangle$. Owing to the orthogonality of the Fock states, the scalar product, as a double sum, reduces to a single sum, as is shown in Eqs. (A.7) and (A.8) for $m \geq n$ and $m \leq n$, respectively. Only those matrix elements are non-vanishing in which $m$ and $n$ have the same parity, so $\tfrac{1}{2}(m-n)$ must be an integer. The finite sums in Eqs. (A.7) and (A.8) are expressed by hypergeometric functions in Eqs. (A.9) and (A.10). Here we display the result in the case $m \geq n$,

$$\langle m|S(\xi)|n\rangle = \exp[-\tfrac{1}{2}(n+\tfrac{1}{2})\eta](\zeta/2)^{(m-n)/2}\sqrt{\frac{m!}{n!}}\frac{1}{\Gamma(\alpha+1)}F\left(-\frac{n}{2},\frac{1-n}{2};\frac{m-n}{2}+1;z\right) \quad (m \geq n), \tag{4}$$

where $z = -\sinh^2|\xi|$ and $\exp(\eta/2) = \cosh|\xi|$. A similar expression comes out in the other case $m \leq n$, as is shown in Eq. (A.10). This formula has been first published by Bargmann (1947), who used the analytic function representation in the Fock-Bargmann space. It was rederived by Popov and Perelomov (1969), by performing the scalar products in the $L^2$ representation, and using generating functions.

The crucial step in our derivation is that now we apply formula 8.962.1 of Gradshteyn and Ryzhik (2000), to connect the hypergeometric functions with the Jacobi polynomials $P_{n/2}^{(\alpha,-1/2)}(x)$, and $P_{(n-1)/2}^{(\alpha,+1/2)}(x)$, where $\alpha = \tfrac{1}{2}(m-n) \geq 0$, and $n/2$ and $(n-1)/2$ are integers (corresponding to even or odd $n$, respectively). This is shown in Eqs. (A.11) and (A.12) in the Appendix. The problem with these expressions is that they contain different Jacobi polynomials with different parameters, depending on the parity. This means that the functional forms of the matrix elements are not the same for even-even and odd-odd transitions. Fortunately, we can get rid off this severe assymmetry by using the formulas 22.5.22 and 22.5.21 of Abramowitz and Stegun (1970) in Eqs. (A.11) and (A.12), to be applied for the even-even and odd-odd transitions, respectively. These formulas, Eqs. (A.13a-b), connect the Jacobi polynomials of the type $P_k^{(\alpha,\pm 1/2)}$ with the Gegenbauer polynomials $C_n^{\alpha+\tfrac{1}{2}}$. In this way we receive a *unifying formula* (in the sub-case $m \geq n$), *valid for both even and odd indeces*, as is expressed in Eq. (A.14). A similar unifying formula (which is also valid for both even and odd paritys) results in the other sub-case $m \leq n$ as is shown in Eq. (A.15). Some additional details of the calculation are explained in Appendix A. On the basis of these considerations, the final result can be summarized in the form

$$\langle m|S(\xi)|n\rangle = e^{-\eta/4}(2e^{i\varphi}\tanh|\xi|)^\alpha \sqrt{\frac{n!}{m!}} \frac{\Gamma(\alpha+\tfrac{1}{2})}{\sqrt{\pi}} C_n^{\alpha+\tfrac{1}{2}}\left(\frac{1}{\cosh|\xi|}\right), \quad \alpha = \tfrac{1}{2}(m-n), \quad (m \geq n), \qquad (5)$$

$$\langle m|S(\xi)|n\rangle = e^{-\eta/4}(-2e^{-i\varphi}\tanh|\xi|)^\lambda \sqrt{\frac{m!}{n!}} \frac{\Gamma(\lambda+\tfrac{1}{2})}{\sqrt{\pi}} C_m^{\lambda+\tfrac{1}{2}}\left(\frac{1}{\cosh|\xi|}\right), \quad \lambda = \tfrac{1}{2}(n-m), \quad (m \leq n). \qquad (6)$$

The formulas in Eq. (5) and Eq. (6) are valid for both even-even and odd-odd transitions, so they are valid in the whole Hilbert space $\mathscr{H} = \mathscr{H}_{1/4} \oplus \mathscr{H}_{3/4}$. The order $\nu$ of the Gegenbauer polynomials $C_n^\nu(x)$ in Eqs. (7) and (8) satisfy the condition $\nu > -1/2$, and this means that the matrix elements are expressed by the classical orthogonal polynomials.

In the simplest special case, when $m = n$, the Gegenbauer polynomials reduce to the Legendre polynomials, according to the relation $C_n^{\tfrac{1}{2}}(x) = P_n(x)$ (see e.g. Gradshteyn and Ryzhik (2000)), i.e.

$$\langle n|S(\xi)|n\rangle = \sqrt{x}P_n(x), \quad x = 1/\cosh r, \quad r = |\xi| \quad (m = n). \qquad (7)$$

In the general case the new formulas for the matrix elements $S_{mn} = \langle m|S(\xi)|n\rangle$, Eqs. (5) and (6), can be brought to another equivalent form, expressed in terms of the associated Legendre functions $P_\mu^\nu(x)$. By using the formula 11.4 (10) of Erdélyi (1953), the Gegenbauer polynomials can be interrelated with the associated Legendre functions,

$$\frac{1}{\sqrt{\pi}}\Gamma(\mu+\tfrac{1}{2})C_{\nu-\mu}^{\mu+\tfrac{1}{2}}(x) = (-1)^\mu 2^{-\mu}(1-x^2)^{-\tfrac{1}{2}\mu} P_\nu^\mu(x),$$

and we have

$$\langle m|S(\xi)|n\rangle = (-1)^k \sqrt{\frac{n!}{m!}} \sqrt{x} P_l^k(x) e^{ik\varphi}, \quad k = \tfrac{1}{2}(m-n), \quad (m \geq n), \qquad (8)$$

$$\langle m|S(\xi)|n\rangle = \sqrt{\frac{m!}{n!}} \sqrt{x} P_l^{k'}(x) e^{-ik'\varphi}, \quad k' = \tfrac{1}{2}(n-m), \quad (m \leq n), \qquad (9)$$

where $x = 1/\cosh|\xi|$ and $l = \tfrac{1}{2}(m+n)$. The transition probability then has the compact form

$$w_{m,n} = |\langle m|S(\xi)|n\rangle|^2 = \frac{n_<!}{n_>!} x |P_{(m+n)/2}^{|m-n|/2}(x)|^2, \qquad (10)$$

where $n_< = \min(m,n)$, $n_> = \max(m,n)$, $x = 1/\cosh r$ (see Popov and Perelomov (1969), Malkin, Man'ko and Trifonov (1970)). The result in Eq. (10) directly shows the symmerty of the transition probalbilities with respect to the interchange of $n$ and $m$. Besides, the matrix elements $S_{mn} = \langle m|S(\xi)|n\rangle$ are formally related to the representation of spatial rotations (see e.g. Aronson, Malkin and Man'ko (1974), Witschel (1975)). By introducing the non-normalized spherical harmonics $X_l^k(\vartheta,\varphi)$,

$$X_l^k(\vartheta,\varphi) = \sqrt{\frac{4\pi}{(2l+1)}} Y_l^k(\vartheta,\varphi) = (-1)^k \left[\frac{(l-k)!}{(l+k)!}\right]^{1/2} P_l^k(\cos\vartheta) e^{ik\varphi}, \qquad (11)$$

Eqs. (8) and (9) can be summarized in a compact formula for the general matrix elements of the squeezing operator. On the basis of Eqs. (8), (9) and (11), we have

$$\langle m|S(\xi)|n\rangle = \sqrt{\cos\theta}\, X_l^k(\theta,\varphi), \quad \cos\theta = 1/\cosh r, \quad l = \tfrac{1}{2}(m+n), \quad k = \tfrac{1}{2}(m-n), \quad (m,n \geq 0), \tag{12}$$

where $0 < \theta < \pi/2$, $l = 0,1,2,\ldots$, $k = -l, -l+1, \ldots, l-1, l$. In the special case $n = 0$, the order equals to $k = l = m/2$, and we encounter with the well-known special formula for spherical harmonics,

$$X_l^l(\theta,\varphi) = (-1)^l \sqrt{\frac{(2l)!}{2^{2l}(l!)^2}} (\sin\theta)^l e^{il\varphi}, \quad l = m/2, \quad \sin\theta = \tanh r. \tag{13}$$

By multiplying this with $\sqrt{\cos\theta}$ we receive the photon number distribution amplitude of a *squeezed vacuum state* $S(\xi)|0\rangle$. We note that, in contrast to our Gegenbauer polynomial expressions, Eqs. (5) and (6), in the other equivalent formulas, given by Eqs. (8), (9) and (12), the degree $l = \tfrac{1}{2}(m+n)$ of the functions is a combination of the initial and final indeces, $n$ and $m$, respectively. This circumstance would severely complicate the calculation for building up the coherent and incoherent superpositions, to be discussed in Section 3.

In the rest of the present section we illustrate some main features of the photon number distribution of a squeezed number state, determined by the Gegenbauer distribution corresponding to Eqs. (5) and (6). Figure 1 shows the distribution for the lowest $n$-values with different parities.

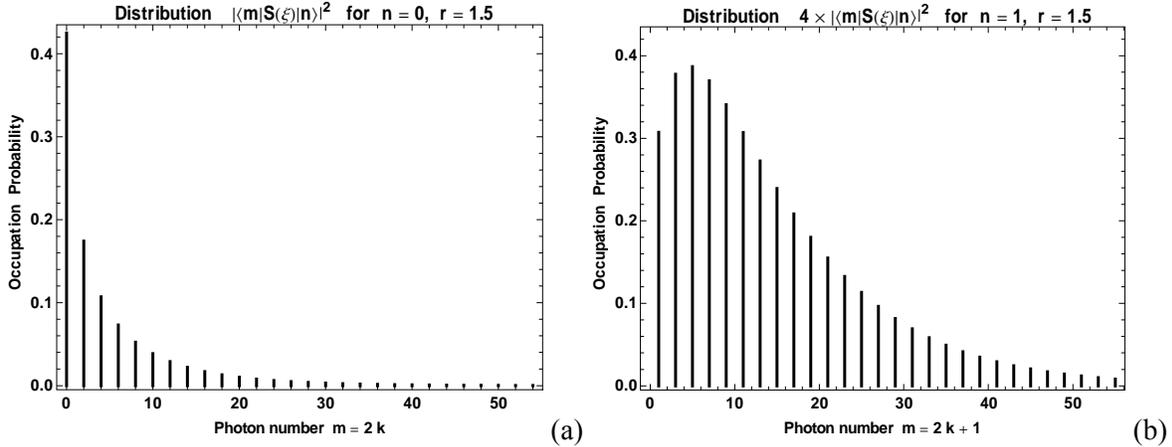

**Figure 1.** Shows the photon number distributions $p_m(n) = |\langle m|S(\xi)|n\rangle|^2$, according to Eqs. (5-6), for the lowest $n$-values with different parities. In both cases $|\xi| \equiv r = 1.5$, and the squeezing parameter $s = e^r = 4.48$. **(a)**: $p_m(0)$ for a squeezed vacuum state $S(\xi)|0\rangle$, in which only even photon number states are occupied; $m = 2k$ (i.e. $m = 0,2,4,\ldots$). This distribution is determined by Eq. (13). The expectation value of the dimensionless energy is $\langle a^+ a + 1/2\rangle = 0.5\cosh 2r = 5.03$. **(b)**: $p_m(1)$ for a squeezed one-photon state $S(\xi)|1\rangle$, in which only odd photon number states are occupied; $m = 2k+1$ (i.e. $m = 1,3,5,\ldots$). The expectation value of the dimensionless energy is $1.5\cosh 2r = 15.1$. For a better comparison with **(a)**, in **(b)** we have plotted $4\times p_m(1)$.

In Fig. 2 the photon number distribution $p_m(5)$ for a squeezed 5-photon state $S(\xi)|5\rangle$ have been plotted, for $|\xi| \equiv r = 1.5$. In this case $n/\cosh^2 r < 1$, and it is justified to use the asymptotic formula 8.936 in Gradshteyn and Ryzhik (2000),

$$\lim_{\nu \to \infty} \nu^{-n/2} C_n^{\nu/2}[t(2/\nu)^{1/2}] = (1/2^{n/2} n!) H_n(t), \qquad (14)$$

which yields $p_m(n) \approx |\psi_n(t_m)|^2 / t_m$, where $\psi_n(x)$ is the usual hermite function for the oscillator eigenfunction in $L^2$ representation, so

$$\langle m|S(\xi)|n\rangle \approx \frac{e^{i\varphi(m-n)/2}}{(\cosh r)\sqrt{t_m}} \psi_n(t_m), \quad \psi_n(x) = \frac{1}{\sqrt{\pi^{1/2} 2^n n!}} H_n(x) \exp(-\tfrac{1}{2} x^2), \quad t_m = \sqrt{\frac{m}{2\cosh^2 r}}. \qquad (15)$$

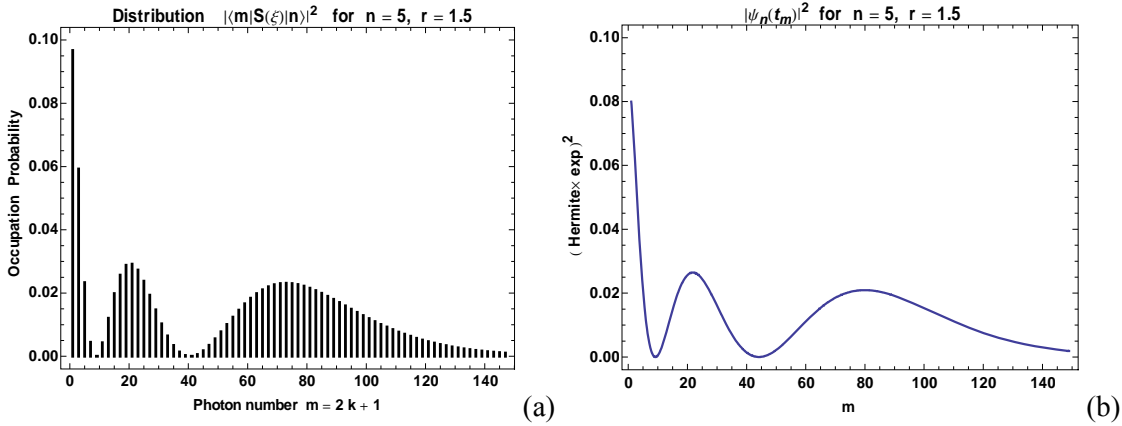

**Figure 2.** Shows the photon number distributions $p_m(5)$, according to Eqs. (5) and (6). In both cases $|\xi| \equiv r = 1.5$, $s = e^r = 4.48$ and $\tanh^2 r = 0.82$. **(a)**: the exact $p_m(5)$ distribution for a squeezed 5-photon state $S(\xi)|5\rangle$, with average energy is $5.5 \cosh 2r = 55.4$. **(b)**: the approximate $p_m(5)$ for the same squeezed 5-photon state. The discrete point are connected for a better visibility of the shape of the distribution Here $n^2 \tanh^2 r = 20.5$ large, but the parameter $n(1 - \tanh^2 r) = 0.9$ is smaller than unity; the distribution is result of that the largely stretched state (in 'coordinate representation') is swept by the narrow original wavefunctions $\psi_m(x)$, whose main contribution comes around the points $t_m$ of Eq. (15).

Fig. 2 (b) has been plotted by using the approximate formula, Eq. (15), for a comparison. The agreement with Fig. 2(a) is quite good. It is very remarkable, that the approximation in Eq. (15), is derived from the exact expressions Eqs. (5) and (6), which are the result of an exclusively abstract, algebraic procedure. So, we would even say, that we 'did not know at the beginning that wave functions in $L^2$ exist, at all.' Still, we find that a quite faithful „coarse-grained picture" of the modulus square of the wave function has been created from the photon number distribution. We note that Popov and Perelomov (1969) have also found this approximation (see their note at the very end of their paper), however they a priori calculated the matrix element in the $L^2$ coordinate representation.

In Fig. 3 the characteristic quantum oscillations are nicely seen in this parameter regime $n/\cosh^2 r \gg 1$. Such oscillation have also been analysed in details by Popov and Perelomov (1969).

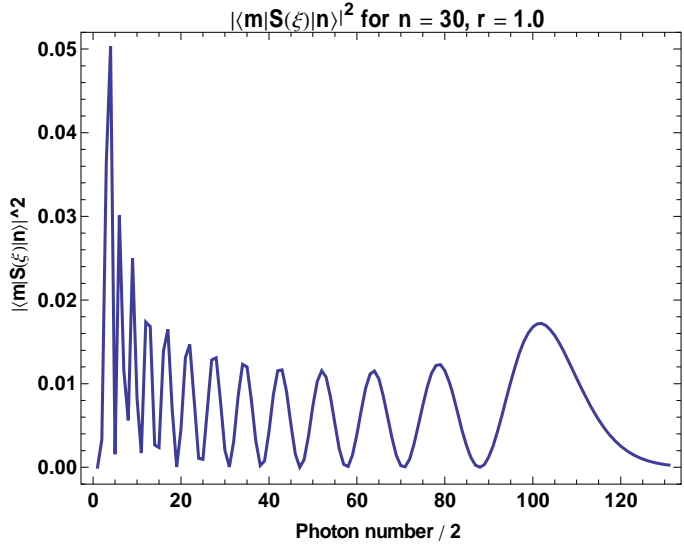

**Figure 3.** Shows the photon number distribution $p_m(30) = |\langle m|S|30\rangle|^2$ for a squeezed 30-photon state $S(\xi)|30\rangle$, with $|\xi| \equiv r = 1$. The discrete point are connected for a better visibiliti of the shape of the distribution. In this regime the quantum oscillations are nicely seen.

### 3. Coherent and incoherent superposition of the matrix elements of squeezing transitions.

In the present Section we determine the coherent and incoherent superpositions of the matrix elements of the squeezing operator, on the basis of the newly found Gegenbauer polynomial expressions in Eqs. (5) and (6). First we note that in the semi-classical description, when the photon field is considered as an external field $F_0 \sin \omega t$, for a free electron in a Volkov state (see e.g. Varró 2021a), the oscillating phase factor $\exp[-i(U_p/4\hbar\omega)\sin 2\omega t]$ appears, which comes from the $A^2$-interaction. $U_p$ is the ponderomotive energy of the electron which is proportional to the intensity of the (laser) field. So, in the semiclassical approximation this factor contributes to the double-photon emission and absorption processes by the Fourier components, governed by the Bessel functions,

$$\exp[-iz\sin 2\omega t] = \sum_{n=-\infty}^{\infty} J_k(z) e^{-i2k\omega \cdot t} \ . \tag{16}$$

The corresponding coherent superposition of the transition amplitudes can be calculated on the basis of the generating formula for the Gegenbauer polynomials (see Gradshteyn and Ryzhik 2000),

$$\Gamma(k+\tfrac{1}{2})\frac{2^k}{\sqrt{\pi}} \sum_{n=0}^{\infty} C_n^{k+\tfrac{1}{2}}(x) \frac{z^n}{\Gamma(2\lambda+n)} = e^{z\cos\theta} \frac{1}{(z\sin\theta)^k} J_k(z\sin\theta) \ .$$

By applying this formula for the (fixed) $2k$-order emission processes, taking Eq. (5), we have

$$\sum_{n=0}^{\infty} \frac{(\beta^*)^{n+2k}}{\sqrt{(n+2k)!}} e^{-\tfrac{1}{2}|\beta|^2} \frac{\alpha^n}{\sqrt{n!}} e^{-\tfrac{1}{2}|\alpha|^2} \langle n+2k|S(\xi)|n\rangle = $$
$$= \exp(-\tfrac{1}{2}|\beta|^2 - \tfrac{1}{2}|\alpha|^2 + \beta^*\alpha\cos\theta) e^{-\eta/4} e^{ik\varphi}\left(\frac{\beta^*}{\alpha}\right)^k J_k(\beta^*\alpha \tanh|\xi|) \tag{17}$$

In Eq. (17) $\alpha$ and $\beta$ are the (optionally highly-excited) initial and final coherent states, which represent the laser field, and the argument of the Bessel function can be related to average photon number, which is connected to the classical intensity. For a correct probability interpretation of this expression, the initial and final coherent states can be considered as von Neumann lattice coherent states (Neumann 1932), as is explained in our recent works in Gombkötő *et al* (2016, 2020), Földi *et al* (2021) and Varró (2021a). By comparing Eq. (17) and the Bessel function amplitudes in Eq. (16), we see that if $\alpha = \beta$, the semiclassical result reproduces the result obtained with the quatized description of the field excitations. However, in the quantal description there is always a chance for $\alpha$ and $\beta$ being different, which expresses the effect of the back-reaction of the electron. So, according to the exact result, Eq. (17), if this back-reaction is not negligible, then the emission (absorption) probability distribution of the $2k$-order processe can be qualitatively different from the semi-classical one.

In the case of the interaction with a spectral component of black-body radiation, or with some chaotic radiation, in the semi-classical description we use an averaging with respect to the field amplitude. In case of a Gaussian distribution we need the average of $J_n^2(z)$ in Eq. (16). We have

$$\frac{1}{I_0}\int_0^\infty dI e^{-I/I_0} J_k^2(\gamma I) = \frac{1}{\pi \gamma I_0} Q_{k-\frac{1}{2}}(1 + 1/2\gamma^2 I_0^2) \quad (k = 0,1,2,\dots), \tag{18}$$

where $I_0$ is the average intensity of the field, and $Q_{k-\frac{1}{2}}(z)$ is a Legendre function of second kind (here it is a so-called toroidal function, see Gradshteyn and Ryzhik, 2000). For the absorption processes we receive the same expression.

We have also calculated the incoheren superpositions of the Gegenbauer amplitudes in Eqs. (5), by using Planck-Bose weights,

$$p_n = (1-b)b^n, \quad b = e^{-h\nu/k_B T} = \frac{\bar{n}}{1+\bar{n}}, \quad p_n = \frac{\bar{n}^n}{(1+\bar{n})^{1+n}}, \quad \bar{n} = \frac{1}{e^{h\nu/k_B T}-1}, \tag{19}$$

where $k_B$ is the Boltzmann constant, and $T$ is the absolute temperature of the black-body radiation. For $2k$-order emission processes we have the result

$$\sum_{n=0}^\infty p_n |\langle n+2k|S(\xi)|n\rangle|^2 = \frac{\cos\theta}{\pi\sqrt{\bar{n}(1+\bar{n})}\sin^2\theta} b^{-k} Q_{k-\frac{1}{2}}(z), \quad z = 1+\frac{1}{2\bar{n}(1+\bar{n})\sin^2\theta}, \quad k = 0,1,2,\dots. \tag{20}$$

For $2k$-order absorptions we have recived the average probability

$$\sum_{n=2l}^\infty p_n |\langle n-2l|S(\xi)|n\rangle|^2 = \frac{\cos\theta}{\pi\sqrt{\bar{n}(1+\bar{n})}\sin^2\theta} b^{+l} Q_{l-\frac{1}{2}}(z), \quad z = 1+\frac{1}{2\bar{n}(1+\bar{n})\sin^2\theta}, \quad l = 0,1,2,\dots. \tag{21}$$

In Eqs. (20) and (21) $\sin^2\theta = \tanh^2 r$. The semi-classical result, Eq. (18), approximates quite well the exact quantum result, calculated on the basis of Eqs. (5) and (6), if the prefactors $b^l = e^{-l \cdot h\nu/k_B T}$ are close to unity, i.e. if $h\nu/k_B T \ll 1$, which just corresponds to the Rayleigh-Jeans limit of the Planck distribution. In the opposite case, in the Wien-limit, when $h\nu/k_B T \gg 1$, the two expressions are quatitatively different.

**4. Summary**

In Section 2 (and in the Appendix) we have proved that matrix elements of the squeezing operator (Bogoliubov transformation) can be expressed in terms of the classical Gegenbauer polynomials. On the basis of this new formula, we have given few illustrative examples for the photon number distribution of a squeezed number state. This new expression makes it possible to determine coherent and incoherent superpositions of these matrix elements in closed analytic forms. In Section 3 we have described multiphoton transitions in the system "charged particle + electromagnetic radiation", induced by a (strong) coherent field or by a spectral component of black-body radiation. The exact results have been compared with the semi-classical ones (the latter are based on non-perturbative matrix elements with c-number radiation fields). It has been found that in the case of interaction with a thermal field, the semi-classical result (calculated with a c-number, Gaussian stochastic field amplitude) yields an acceptable approximation only in the Rayleigh-Jeans limit (hν/kT<<1). It has been explicitly shown that in the Wien limit (hν/kT>>1) the semiclassical formula for the multiphoton absorption and emission probabilities cannot even 'mimic' its quantum counterpart, because it has a different functional form.

**Acknowledgments**
The author thanks Prof. Dr. Wolfgang Schleich of Universität Ulm for the illuminating discussions on the subject of the present paper, we had during his recent visit in Budapest. Support by the ELI-ALPS project is acknowledged. The ELI-ALPS Project No. GINOP 2.3.6-15 is supported by the European Union and co-financed by the European Regional Development Fund.

**Appendix A. Algebraic derivation of the matrix elements of the squeezing operator between photon number eigenstates.**

In the present Appendix A we give the details of the calculation of the matrix elements between the number eigenstates of the unitary squeezing operator, $S_{m,n} = \langle m|S(\xi)|n\rangle$, where

$$S(\xi) = \exp[\tfrac{1}{2}\xi(a^+)^2 - \tfrac{1}{2}\xi^*(a)^2], \quad S(\xi) = \exp(\xi K_+ - \xi^* K_-), \quad \xi = |\xi|e^{i\varphi}, \tag{A.1}$$

with $K_+ = \tfrac{1}{2}(a^+)^2$ and $K_- = \tfrac{1}{2}(a)^2$. By introducing $K_0 = \tfrac{1}{4}(aa^+ + a^+a)$, we have the closed set of generators of a special Lie algebra of the $SU(1,1)$ group, having the commutation relations $[K_0, K_\pm] = \pm K_\pm$, and $[K_-, K_+] = 2K_0$. The Casimir operator $\hat{C}_2 = K_0^2 - \tfrac{1}{2}(K_+K_- + K_-K_+)$ commutes with all the generators, hence it is proportional with the unit operator, $C_2 = \kappa(\kappa-1)\hat{I}$, where $\kappa$ is called the Bargmann index. In the special case under discussion, we have $C_2 = -(3/16)\hat{I}$, which yields $\kappa = 1/4$ or $\kappa = 3/4$. For $\kappa = 1/4$, the states $|n\rangle$ with even $n$ form a basis of the unitary irreducible representation space of the group $SU(1,1)$, and the states $|n\rangle$ with odd $n$ form a basis corresponding to $\kappa = 3/4$. In the 'even representation space' $K_0|2k\rangle = (k+1/4)|2k\rangle$ ($k = 0,1,2,...$), and in the 'odd representation space' $K_0|2l+1\rangle = (l+3/4)|2l+1\rangle$ ($l = 0,1,2,...$). Each equation can be written in the form $K_0|\psi_\nu\rangle = \nu|\psi_\nu\rangle$, where $\nu - \kappa = 0,1,2,...$. Of course, the complete Hilbert space is the direct sum of these sub-spaces, $\mathscr{H} = \mathscr{H}_{1/4} \oplus \mathscr{H}_{3/4}$. In the present and in similar contexts, the properties of the $SU(1,1)$ group have been used by Popov and Perelomov (1969), Malkin, Man'ko and Trifonov (1970), Gilmore (1974), Perelomov (1977), Kelemen (1975), Witschel (1975), Fisher, Nieto and Sandberg (1984), Schumaker and Caves (1985), Varró (2015), Wünsche (1999, 2003, 2017a-b).

Since neither the commutation relations, nor the parameters depend on the Bargmann index, the ordering (factorization) of $S(\xi)$ can be performed in any concrete representation, of course. Perhaps the simplest way to derive the normal (and anti-normal) form of $S(\xi)$ is to use the 'spinor representation' of the generators (see e.g. Gilmore (1974), Fisher, Nieto and Sandberg (1984) or Schumaker and Caves (1985)). It is an easy matter to check that the following $2\times 2$ matrices

$$-\sigma_+ = \begin{pmatrix} 0 & -1 \\ 0 & 0 \end{pmatrix}, \quad \sigma_- = \begin{pmatrix} 0 & 0 \\ 1 & 0 \end{pmatrix}, \quad \tfrac{1}{2}\sigma_3 = \tfrac{1}{2}\begin{pmatrix} 1 & 0 \\ 0 & -1 \end{pmatrix}$$

satisfy the same commutation rules as $K_+$ and $K_-$ and $K_0$, respectively, thus $-\sigma_+$, $\sigma_-$ and $\tfrac{1}{3}\sigma_3$ also form an $SU(1,1)$ algebra. The ordering can be performed by expanding the exponential expressions of the $2\times 2$ matrix representants (which are also $2\times 2$ matrices), and comparing the parameters of the original exponential expression with the parameters of the new exponential factors. From this we can determine the coefficients of $(-\sigma_+)$, $\sigma_-$ and $\tfrac{1}{2}\sigma_3$ in the new exponential factors, and these coefficients must be identified with that of $K_+$ and $K_-$ and $K_0$ in the normal (anti-normal) form. In the 'spinor representation' we have

$$S(\xi) = \exp(\xi K_+ - \xi^* K_-) \rightarrow \exp\left[-\begin{pmatrix} 0 & \xi \\ \xi^* & 0 \end{pmatrix}\right] = \begin{pmatrix} \cosh|\xi| & -e^{i\varphi}\sinh|\xi| \\ -e^{-i\varphi}\sinh|\xi| & \cosh|\xi| \end{pmatrix}, \tag{A.2}$$

$$\exp(\zeta K_+)\exp(-\eta K_0)\exp(\zeta' K_-) \rightarrow \begin{pmatrix} 1 & -\zeta \\ 0 & 1 \end{pmatrix}\begin{pmatrix} e^{-\eta/2} & 0 \\ 0 & e^{\eta/2} \end{pmatrix}\begin{pmatrix} 1 & 0 \\ \zeta' & 1 \end{pmatrix} = \begin{pmatrix} e^{-\eta/2} - \zeta\zeta' e^{\eta/2} & -\zeta e^{\eta/2} \\ e^{\eta/2}\zeta' & e^{\eta/2} \end{pmatrix}. \tag{A.3}$$

By equating the components of the right hand sides,
$\cosh|\xi| = e^{\eta/2}$, i.e. $\eta = 2\log\cosh|\xi|$, $\quad -e^{i\varphi}\sinh|\xi| = -\zeta e^{\eta/2} = -\zeta\cosh|\xi|$, i.e. $\zeta = e^{i\varphi}\tanh|\xi|$,
$-e^{-i\varphi}\sinh|\xi| = \zeta' e^{\eta/2} = \zeta'\cosh|\xi|$, i.e. $\zeta' = -e^{-i\varphi}\tanh|\xi| = -\zeta^*$. (A.4)

Thus, the normal form of $S(\xi)$ becomes

$$S(\xi) = \exp(\xi K_+ - \xi^* K_-) = \exp(\zeta K_+)\exp(-\eta K_0)\exp(-\zeta^* K_-). \tag{A.5}$$

On the basis of (A.4), the new parameters are determined to be $\eta = 2\log\cosh|\xi|$, $\zeta = e^{i\varphi}\tanh|\xi|$. We note that the anti-normal form can also be obtained similarly,

$$\hat{S}(\xi) = \exp(\xi K_+ - \xi^* K_-) = \exp(-\zeta^* K_-)\exp(\eta K_0)\exp(\zeta K_+). \tag{A.6}$$

With the help of the normal form in (A.5) the calculation of the matrix element $\langle m|S(\xi)|n\rangle$ is now straightforward. The effect of $\exp(-\zeta^* \tfrac{1}{2}a^2)$ on $|n\rangle$ yields a finite sum, in which the highest power of $-\zeta^*$ is $[n/2]$, where $[x]$ denotes the integer part of $x$ (i.e. the smallest integer, which is larger or equal to $x$). The factor in the middle of the expression in (A.5) is diagonal, i.e. $\exp(-\eta K_0)|k\rangle = \exp[-(\eta/2)(k+\tfrac{1}{2})]|k\rangle$ for any $k$. The factor $\langle m|\exp(\zeta K_+) = (\exp(\zeta^* \tfrac{1}{2}a^2)|m\rangle)^+$ on the left yields also a finite sum, in which the highest power of $\zeta$ is $[m/2]$. The scalar product of the two sums terminates at the smaller maximum summation index $\min([m/2],[n/2])$. Any term like $\langle m|(a^{+2})^l(a^2)^k|n\rangle$ is proportional to $\langle m|n-2k+2l\rangle = \delta_{m,n-2k+2l}$, thus, only those matrix elements are non-vanishing in which $m$ and $n$ have the same parity, and this means that their difference $m-n$ must be an even number, $m-n = 2\alpha$, where $\alpha$ is an integer. By taking all these considerations into account, a straightforward calculation leads to the explicit result,

$$\langle m|S(\xi)|n\rangle = \exp[-\tfrac{1}{2}(n+\tfrac{1}{2})\eta](\zeta/2)^{(m-n)/2}\sqrt{m!}\sqrt{n!}\sum_{k=0}^{[n/2]}\frac{(-|\zeta|^2 e^\eta/4)^k}{[\tfrac{1}{2}(m-n)+k]!k!(n-2k)!}, \quad (m\geq n), \quad (A.7)$$

$$\langle m|S(\xi)|n\rangle = \exp[-\tfrac{1}{2}(m+\tfrac{1}{2})\eta](-\zeta^*/2)^{(n-m)/2}\sqrt{m!}\sqrt{n!}\sum_{l=0}^{[m/2]}\frac{(-|\zeta|^2 e^\eta/4)^l}{[\tfrac{1}{2}(n-m)+l]!l!(m-2l)!} \quad (m\leq n). \quad (A.8)$$

These expressions (A.7) and (A.8) are equivalent to the ones derived long ago by Husimi (1953), who used two-variable generating functions of $S_{mn}$, and the $L^2$ representation (Hermite functions) for the oscillator basis states (see also Popov and Perelomov 1969). For a real $\zeta$ Eq. (A.7) reduces to the equation (27) in Kelemen's (1975) paper. See also Tanabe (1973), Rashid (1975), Satyanarayana (1985) and Mendaš and Popović (1995).

*Explicit form of the matrix elements of the squeezing operator between photon number eigenstates, expressed in terms of Gegenbauer polynomials.*

Now, our task is to express the finite sums (A.7) and (A.8) in terms of known functions, which makes it possible to perform the analytic calculations shown in the main text. Let us consider the sum on the right hand side of (A.7), which refers to the sub-case $m\geq n$. We observe that the first factorial in the denominator can be expressed as

$$[\tfrac{1}{2}(m-n)+k]! = [\tfrac{1}{2}(m-n)]!\big(\tfrac{1}{2}(m-n)+1\big)_k, \quad (a)_k = (a)\cdot(a+1)\cdots(a+k-1) = \frac{\Gamma(a+k)}{\Gamma(a)},$$

where we have introduced Pochhammer's symbol $(a)_k$, which can also be expressed in terms of the gamma functions (Abramowitz and Stegun, 1970). Moreover, from the explicit form of the factorials, one can show

$$\frac{n!}{(n-2k)!} = 2^{2k}\left(-\frac{n}{2}\right)_k\left(\frac{1-n}{2}\right)_k.$$

These manipulations allow us to bring into consideration the hypergeometric series $F(a,b;c;z)$, which may be used to represent orthogonal polynomials,

$$F(a,b;c;z) = 1 + \frac{a\cdot b}{c\cdot 1}z + \frac{a(a+1)b(b+1)}{c(c+1)\cdot 1\cdot 2}z^2 + \ldots = \sum_{k=0}^\infty \frac{(a)_k(b)_k}{k!(c)_k}z^k.$$

In our formula (A.7) this series terminates when the summation index takes on the value $k=[n/2]$ (and in (A.8) the sum terminates at $k=[m/2]$). The expressions (A.7) and (A.8) for the matrix elements can then be brought to the equivalent forms, expressed in terms of the hypergeometric functions,

$$\langle m|S(\xi)|n\rangle = \exp[-\tfrac{1}{2}(n+\tfrac{1}{2})\eta](\zeta/2)^{(m-n)/2}\sqrt{\frac{m!}{n!}}\frac{1}{\Gamma(\alpha+1)}F\left(-\frac{n}{2},\frac{1-n}{2};\frac{m-n}{2}+1;z\right) \quad (m\geq n), \quad (A.9)$$

$$\langle m|S(\xi)|n\rangle = \exp[-\tfrac{1}{2}(m+\tfrac{1}{2})\eta](-\zeta^*/2)^{(n-m)/2}\sqrt{\frac{n!}{m!}}\frac{1}{\Gamma(\lambda+1)}F\left(-\frac{m}{2},\frac{1-m}{2};\frac{n-m}{2}+1;z\right) \quad (m\leq n), \quad (A.10)$$

where $z = -|\zeta|^2 e^\eta = -\sinh^2|\xi|$, and $\alpha = \tfrac{1}{2}(m-n)$ in (A.9) and $\lambda = \tfrac{1}{2}(n-m) = |\alpha|$ in (A.10) are non-negative integers. It can be seen that Equations (A.9) and (A.10) go over into each other if we exchange $n$ and $m$, take the complex conjugate of the resulting expression, and multiply this by $(-1)^{(m-n)/2}$. Thus, it is enough to detail the forthcoming derivation only for one of these equations, e.g. (A.9). In addition we note that Eqs. (A.9) and (A.10) are equivalent to equations (10.28a-b) of Bargmann (1947) (see also Sciarrino and Toller (1967)).

Consider the transitions between *even* photon number states, i.e., assume that $n=2k$ where $k=0,1,2\ldots$ (and $m=2q$ with $q=0,1,2\ldots$). Then, on the right hand side of equation (A.9) we have $F(-k,-k-\beta;\alpha+1;z)$, where $\beta = -1/2$ and $\alpha = \tfrac{1}{2}(m-n) = k-q$ is the same non-negative integer, as has been defined above. We apply the formula 8.962.1 of Gradshteyn and Ryzhik (2000), p. 990, which gives the connection between the Jacobi polynomials and the hypergeometric functions,

$$P_k^{(\alpha,\beta)}(x) = \frac{\Gamma(k+\alpha+1)}{k!\Gamma(\alpha+1)}\left(\frac{x+1}{2}\right)^k F\left(-k,-k-\beta;\alpha+1;\frac{x-1}{x+1}\right), \quad \frac{x-1}{x+1} = z, \tag{*}$$

where $P_k^{(\alpha,\beta)}(x)$ is a Jacobi polynomial of degree $k = n/2$, and $\beta = -1/2$ in the present case. The argument of the Jacobi polynomial can simply be expressed as $x = (1+z)/(1-z)$, where $z = -\sinh^2|\xi|$. By taking this connection (*) into account in Equation (A.9), we receive

$$\langle m|S(\xi)|n\rangle = e^{-\eta/4}(\zeta/2)^\alpha \frac{\sqrt{m!}}{\sqrt{n!}} \frac{(\tfrac{1}{2}n)!}{(\tfrac{1}{2}m)!} P_{n/2}^{(\alpha,-1/2)}(x), \quad x = \frac{1+z}{1-z}, \tag{A.11}$$

($m \geq n$, even $m$ and even $n \geq 0$).

Consider now the transitions between *odd* photon number states, i.e., assume that $n = 2k+1$ where $k = 0,1,2\ldots$ (and $m = 2q+1$ with $q = 0,1,2\ldots$). We take into account that the hypergeometric functions are symmetric in the first two parameters; $F(a,b;c;z) = F(b,a;c;z)$. Then, on the right hand side of equation (A.9) we have $F(-k,-k-\beta';\alpha+1;z)$, but now $\beta' = +1/2$. As before, we again apply the same formula 8.962.1 of Gradshteyn and Ryzhik (2000), denoted by (*) above, for the connection of the Jacobi polynomials and the hypergeometric functions, and receive

$$\langle m|S(\xi)|n\rangle = e^{-\eta/4}(\zeta/2)^\alpha \frac{\sqrt{m!}}{\sqrt{n!}} \frac{[\tfrac{1}{2}(n-1)]!}{[\tfrac{1}{2}(m-1)]!} \sqrt{\frac{1+x}{2}} P_{(n-1)/2}^{(\alpha,+1/2)}(x), \quad x = \frac{1+z}{1-z}, \tag{A.12}$$

($m \geq n$, odd $m$ and odd $n \geq 1$).

The argument of the Jacobi polynomial is $x = (1+z)/(1-z)$ (where $z = -\sinh^2|\xi|$), and the other parameters have been defined in (A.4); $\zeta = e^{i\varphi}\tanh|\xi|$, and $e^{\eta/2} = \cosh|\xi|$.

Finally, we use the formulas 22.5.22 and 22.5.21 of Abramowitz and Stegun (1970) in Eq. (A.11) and Eq. (A.12), referring to the even-even and odd-odd transitions, respectively,

$$P_k^{(\alpha,-1/2)}(x) = \frac{(\tfrac{1}{2})_k}{(\alpha+\tfrac{1}{2})_k} C_{2k}^{\alpha+\tfrac{1}{2}}\left(\sqrt{\frac{1+x}{2}}\right), \quad P_k^{(\alpha,+1/2)}(x) = \frac{(\tfrac{1}{2})_{k+1}}{(\alpha+\tfrac{1}{2})_{k+1}\sqrt{\frac{1+x}{2}}} C_{2k+1}^{\alpha+\tfrac{1}{2}}\left(\sqrt{\frac{1+x}{2}}\right), \tag{A.13a}$$

$$x = \frac{1+z}{1-z}, \quad \frac{1+x}{2} = \frac{1}{1-z}, \quad z = -\sinh^2|\xi|, \quad \sqrt{\frac{1+x}{2}} = \frac{1}{\cosh|\xi|}. \tag{A.13b}$$

Here $C_n^\nu(x)$ are the Gegenbauer polynomials (see e.g. Gradshteyn and Ryzhik (2000)). Inserting the first and the second equations of (A.13a) in Eqs. (A.11) and (A.12), respectively, and taking (A.13b) into account, we receive a *unifying formula* (in the sub-case $m \geq n$), *valid for both even and odd indeces*,

$$\langle m|S(\xi)|n\rangle = e^{-\eta/4}(2e^{i\varphi}\tanh|\xi|)^\alpha \sqrt{\frac{n!}{m!}} \frac{\Gamma(\alpha+\tfrac{1}{2})}{\sqrt{\pi}} C_n^{\alpha+\tfrac{1}{2}}\left(\frac{1}{\cosh|\xi|}\right), \quad \alpha = \tfrac{1}{2}(m-n), \quad (m \geq n). \tag{A.14}$$

A similar unifying formula (which is also valid for both parityies) results in the other sub-case $m \leq n$,

$$\langle m|S(\xi)|n\rangle = e^{-\eta/4}(-2e^{-i\varphi}\tanh|\xi|)^\lambda \sqrt{\frac{m!}{n!}} \frac{\Gamma(\lambda+\tfrac{1}{2})}{\sqrt{\pi}} C_m^{\lambda+\tfrac{1}{2}}\left(\frac{1}{\cosh|\xi|}\right), \quad \lambda = \tfrac{1}{2}(n-m), \quad (m \leq n). \tag{A.15}$$

In deriving Equation (A.14) from Eqs. (A.11) and (A.12), we have also taken into account the so-called doubling formula $\Gamma(2x) = \pi^{-1/2} 2^{2x-1}\Gamma(x)\Gamma(x+\tfrac{1}{2})$ of the gamma function (see e.g. Gradshteyn and Ryzhik 2000), and the explicit form of the parameter $\eta = 2\log\cosh|\xi|$. According to the functional relation $C_n^{\frac{1}{2}}(x) = P_n(x)$, if $m = n$, then, both in (A.14) and (A.15), the Gegenbauer polynomials reduce to the Legendre polynomial $P_n(x)$,

$$\langle n|S(\xi)|n\rangle = \sqrt{x}P_n(x), \quad x = 1/\cosh r, \quad r = |\xi| \qquad (m = n). \qquad (A.16)$$

We also note that, in contrast to Eqs. (A.11) and (A.12) being separately valid in the sub-spaces $\mathscr{H}_{1/4}$ and $\mathscr{H}_{3/4}$, respectively, the unifying formulas (A.14) (and also (A.15)) are valid in the whole Hilbert space $\mathscr{H} = \mathscr{H}_{1/4} \oplus \mathscr{H}_{3/4}$. On "unifying formula" we mean that the functional form of the expressions are the same for both the $\mathscr{H}_{1/4}$ (even) and the $\mathscr{H}_{3/4}$ (odd) subspaces. This is not the case for Eqs. (A.11) and (A.12), because they contain different Jacobi polynomials, $P_{n/2}^{(\alpha,-1/2)}(x)$ and $P_{(n-1)/2}^{(\alpha,+1/2)}(x)$, with different parameters, respectively.

**References.**